\documentclass[conference, compsoc]{IEEEtran}
\IEEEoverridecommandlockouts
\usepackage[hidelinks]{hyperref}
\usepackage{cite}
\usepackage{amsmath,amssymb,amsfonts}
\usepackage{algorithmic}
\usepackage{graphicx}
\usepackage{textcomp}
\usepackage{xcolor}
\usepackage{tabularx}
\usepackage{booktabs}
\usepackage{multirow}
\usepackage{float}
\usepackage{fancyhdr}
\usepackage[numbers]{natbib}
\usepackage[ruled,vlined]{algorithm2e}
\usepackage{xcolor}
\usepackage[symbol]{footmisc}
\usepackage{enumitem}

\setlist[itemize]{leftmargin=*}

\def\BibTeX{{\rm B\kern-.05em{\sc i\kern-.025em b}\kern-.08em
    T\kern-.1667em\lower.7ex\hbox{E}\kern-.125emX}}
\begin{document}

\title{Energy-efficient Federated Learning with Dynamic Model Size Allocation\\
}

\makeatletter
\newcommand{\linebreakand}{%
  \end{@IEEEauthorhalign}
  \hfill\mbox{}\par
  \mbox{}\hfill\begin{@IEEEauthorhalign}
}
\makeatother

\author{
\IEEEauthorblockN{M S Chaitanya Kumar *}
\IEEEauthorblockA{\textit{Computer \& Information Sciences} \\
\textit{University of Hyderabad}\\
TG, India \\
20mcme20@uohyd.ac.in}
\and
\IEEEauthorblockN{Sai Satya Narayana J *}
\IEEEauthorblockA{\textit{Computing Technologies} \\
\textit{SRM University}\\
Chennai, TN, India \\
jj1039@srmist.ac.in}
\and
\IEEEauthorblockN{Yunkai Bao}
\IEEEauthorblockA{\textit{Electrical and Software Engineering} \\
\textit{University of Calgary}\\
Calgary, AB, Canada \\
yunkai.bao@ucalgary.ca}
\linebreakand
\IEEEauthorblockN{Xin Wang}
\IEEEauthorblockA{\textit{Geomatics Engineering} \\
\textit{University of Calgary}\\
Calgary, AB, Canada \\
xcwang@ucalgary.ca}
\and
\IEEEauthorblockN{Steve Drew}
\IEEEauthorblockA{\textit{Electrical and Software Engineering} \\
\textit{University of Calgary}\\
Calgary, AB, Canada \\
steve.drew@ucalgary.ca}

}

\maketitle
\author{}
\begin{abstract}
Federated Learning (FL) presents a paradigm shift towards distributed model training across isolated data repositories or edge devices without explicit data sharing. Despite of its advantages, FL is inherently less efficient than centralized training models, leading to increased energy consumption and, consequently, higher carbon emissions. In this paper, we propose CAMA, a carbon-aware FL framework, promoting the operation on renewable excess energy and spare computing capacity, aiming to minimize operational carbon emissions. CAMA introduces a dynamic model adaptation strategy which adapts the model sizes based on the availability of energy and computing resources. Ordered dropout is integratged to enable the aggregation with varying model sizes. Empirical evaluations on real-world energy and load traces demonstrate that our method achieves faster convergence and ensures equitable client participation, while scaling efficiently to handle large numbers of clients. The source code of CAMA is available at \url{https://github.com/denoslab/CAMA}.

\end{abstract}

\footnotetext[1]{Work done during the Mitacs Globalink Research Internship (GRI) at the University of Calgary in Canada.}


\section{Introduction}
The massive increase in the number of edge devices has called for routine storage of application data, formulating edge computing clusters with great potential for mining their knowledge. 
The privacy concerns of the data being generated at edge devices and the increase in computational capabilities of edge devices have been coupled with a significant paradigm shift from traditional machine learning (ML). This paradigm shift is termed as Federated Learning (FL), a distributed ML framework that enables several clients to produce a global inference model by aggregating locally trained model parameters without sharing local data \cite{b1,fed}. 


\begin{figure}[tb]
\centerline{\includegraphics[trim={0 4cm 0 4cm},clip,scale = 0.3]{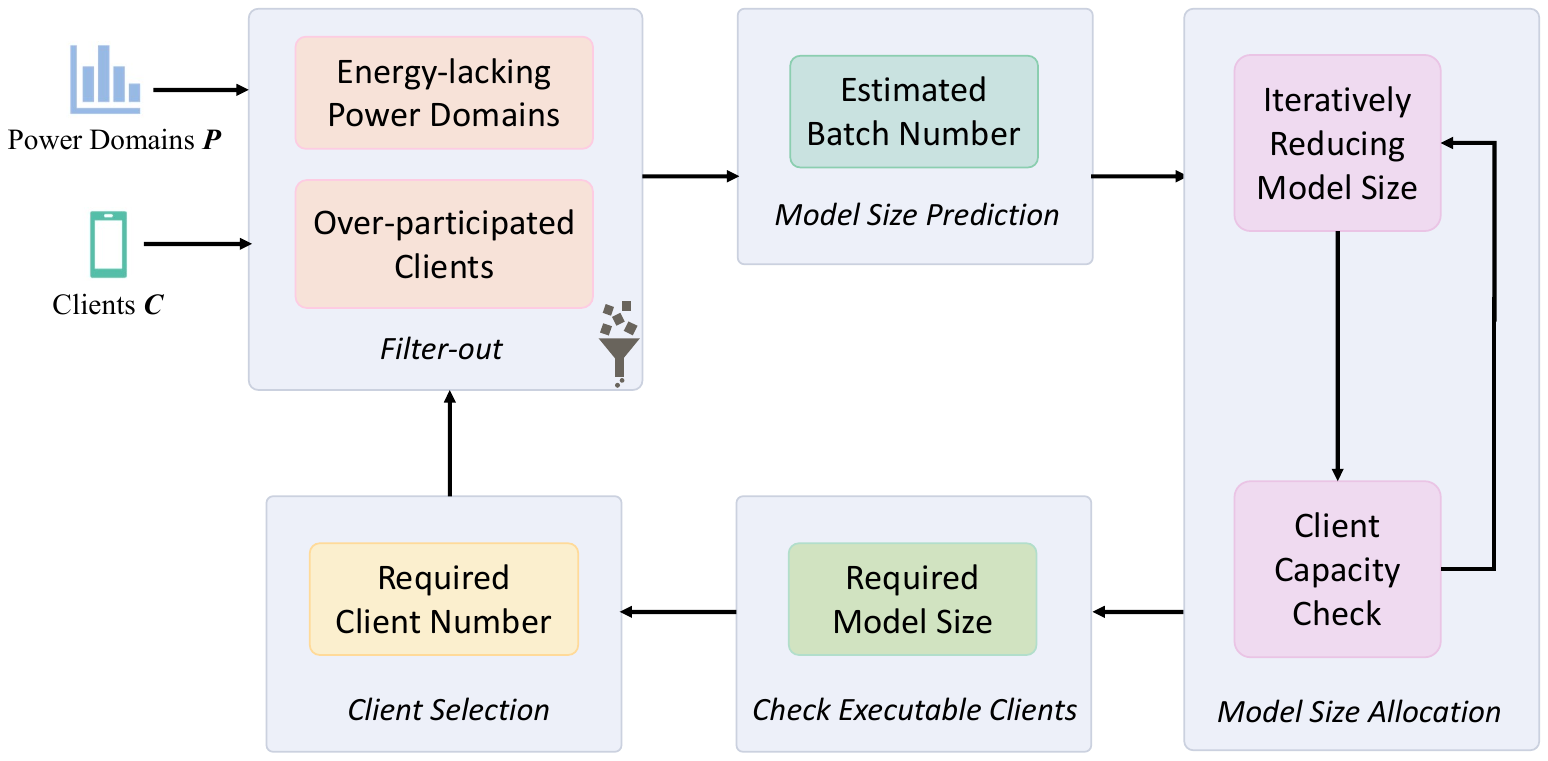}}
\caption{ This figure illustrates the overall framework of CAMA algorithm. In each iteration, the server excludes power domains that have insufficient energy, and clients with too high participation frequency. Then it estimates the number of batches that the clients can handle and dynamically divides the model size for them. Finally, the server samples the clients that meet the requirements.}
\label{fig:intro}
\end{figure}

FL approaches require considerably more
training rounds than traditional ML and are often executed on infrastructure that is less energy-efficient than centralized GPU clusters. As a result, there is a notable rise in overall energy use and related emissions \cite{b3,b3_1}.
In addition, training a state-of-the-art model now requires substantial computational resources, which demand considerable energy, along with the associated financial and environmental costs. 

To address the limitations of current FL systems in terms of energy efficiency and computational resource optimization, we introduce Carbon Aware Model Adaption (CAMA) in FL, designed explicitly for heterogeneous and geo-distributed environments. The core contributions of our proposal are outlined as follows:
\begin{itemize}
    \item We present a novel system design that extends the FedZero framework \cite{b3} by incorporating dynamic adaptation of model sizes \cite{b1} based on renewable energy forecasts and spare computing capacities \cite{b3}. This adaptation allows for more efficient use of computational resources, aligning clients' computational tasks with available green energy sources and ensuring that FL training are executed within shared energy budgets at runtime.
    \item We propose an ordered dropout FL aggregation strategy. This strategy enables the aggregation of models of varying sizes without compromising the learning process \cite{dropout,b1}. This strategy facilitates fast convergence and equitable client participation, even under fluctuating energy and resource availability.
    \item CAMA has been evaluated on various datasets, models, and scenarios. Numerical results demonstrate that CAMA enables faster and more energy-efficient training. 
\end{itemize}

\section{Problem Statement}
Our aim is to train a global model using a federated learning setup that leverages renewable excess energy and spare computing capacity of clients. In this approach we aim to achieve faster convergence even with highly non-IID data, all while preserving privacy or even while sharing the batch normalization layers in intermediate rounds.

\subsection{Client Selection Strategy}

\subsubsection{Client Registration}
In alignment with the framework established by FedZero \cite{b3}, clients are required to register specific metrics with the server prior to participation. These metrics include the energy consumption per batch and the control plane address, which indicates the client's power domain. 
This registration process enables the server to estimate each client's resource consumption effectively.

\subsubsection{Fairness of Participation}
Our strategy emphasizes the importance of fairness in client participation. Clients that contribute significantly to model convergence, either through the quality or uniqueness of their data, are assigned higher probabilities of being selected in each training round. 

\subsubsection{Calculation of Selection Probabilities}
The determination of a client's probability of selection in any given round is influenced by multiple factors, designed to promote inclusivity and fairness:
\begin{itemize}
    \item \textbf{Exclusion After Participation:} To prevent the dominance of certain clients in the training process, a client that has participated in the last round is temporarily excluded from selection for a predefined number of subsequent rounds. This mechanism effectively reduces their participation to zero during this exclusion period, allowing other clients to contribute.
    \item \textbf{Participation Frequency:} The number of rounds a client has previously participated in is considered, ensuring that all clients have equitable opportunities to contribute over time.
    \item \textbf{Model Size Contribution:} The size of the model with which a client has participated in previous rounds is also a factor, recognizing the varying computational efforts contributed by different clients.
    \item \textbf{Average Participation Rate:} The overall engagement of clients is measured by the average number of rounds participated in by all clients, providing a benchmark for assessing individual client contributions.
\end{itemize}

By incorporating these factors into the client selection process, our approach aims to balance resource optimization with the need for fairness and diversity in model training. This strategy not only enhances the efficiency and sustainability of federated learning systems but also ensures that the resulting models are robust, fair, and representative of the collective data landscape.

The overall probability of clients being selected in round $r_i$ compared to the one given in FedZero \cite{b3} is given by Eq. \ref{eq01}, and the statistical utility calculation of a client is the same as is defined in Oort \cite{b4} and FedZero \cite{b3}:

\begin{equation}
    P(c) = 
    \begin{cases}
        \frac{1}{(wp(c) - \omega)^{\alpha}}, & \text{if } wp(c) - \omega \geq 1 \\
        1, & \text{otherwise}
    \end{cases}
\label{eq01}
\end{equation}

$wp(c)$ represents the weighted participation count of the client 
$c$. This ensures that clients who have participated with larger model sizes will have a lower probability of being selected in subsequent rounds compared to clients who have participated with smaller model sizes.
$\omega$ denotes the mean of weighted participated rounds of all clients $\omega = mean \{wp(c) \forall c \in C\}$. $ \alpha $ is the parameter as defined in fedzero\cite{b3}, which is user-defined and generally describes the rate at which a client's probability decreases based on the difference between the mean of the weighted participation rounds of all clients and the weighted participation of that particular client.


For faster convergence in a much varied data distribution among clients, we filter out the clients based on the statistical utility similar to those defined in Oort \cite{b4} and FedZero \cite{b3}. The statistical utility calculation of a client is given by:
\begin{equation}
    \sigma_c = 
    \begin{cases} 
        |B_c| \sqrt{\frac{1}{|B_c|} \sum_{k \in B_c} \text{loss}(k)^2}, & \text{if } p(c) \geq 1 \\
        1, & \text{otherwise}
    \end{cases}
\end{equation}
Oort approximates statistical client utility based on the number of
available training samples $B_c$ and the local training loss, which is expected to correlate with the gradient norm \cite{b3}. Therefore, the modified client selection algorithm from FedZero, described in Algorithm~\ref{alg:clientSelection}, is utilized in CAMA. In each iteration, the server will first filter out power domains without excess energy (Line 4) and clients that over-participated in the past (Line 5). Then, each power domain will make prediction about the model size it can handle based on the estimated number of batches it is expected to process (Line 6-8). After estimation, the number of clients that could execute with model size = 1 is checked (Line 9). Eventually, $n$ clients are selected so that the proportion of clients with each model size remains nearly the same (Line 10).

\renewcommand{\algorithmicrequire}{\textbf{Input:}}
\renewcommand{\algorithmicensure}{\textbf{Output:}}

\begin{algorithm}
\caption{Client Selection Strategy}\label{alg:clientSelection}
\begin{algorithmic}[1]
    \REQUIRE $C$: set of clients, $P$: set of power domains, $n$: minimum number of clients to select 
    \ENSURE Selected clients
    
    \STATE $i \gets 0$
    \STATE $\text{count}_1 \gets 0$
    \REPEAT
        \STATE $\bar{P} \gets \{\forall p \in P, \forall t = 1, ..., i : r_{p,t} > 0\}$
        \STATE $\bar{C} \gets \{\forall c \in C : \sigma_{c} > 0\}$
        \FOR{$p \in \bar{P}$}
            \STATE $\bar{C} \gets \{(c, $\text{model\_size}$(\sum_{t=0}^d \min(m_{c,t}^{\text{spare}}, \frac{r_{p,t}}{\delta_{c}}), c) )\mid  c \in C_p\}$
        \ENDFOR
        \STATE $\text{count}_1 \gets |\{ c \in C' \mid c = 1 \}|$
        \STATE $\text{clients} \gets \text{sort\_select}(\bar{C}, n)$
        \STATE $i \gets i + 1$
    \UNTIL{$|clients| > n \text{ and } {count}_1 > 2$}
    \RETURN $\text{clients}$
\end{algorithmic}
\end{algorithm}

\begin{algorithm}
\caption{Determine Model Size Based on Batches}\label{alg:determineModelSize}
\begin{algorithmic}[1]
    \REQUIRE $batches$, $client$
    \ENSURE $model\_size$
    \STATE $mr \gets 1$
    \STATE $b_c \gets dataset\_batches(client) \times epochs$
    \FOR{$i \gets 1$ \KwTo $5$}
        \IF{$batches \geq b_c \times mr$}
            \RETURN $mr$
        \ENDIF
        \STATE $mr \gets mr / 2$
    \ENDFOR
    \RETURN Default size $\mu$
\end{algorithmic}
\end{algorithm}

\subsection{Dynamic Model Size Allocation Based on Batch Processing Capabilities}

As proposed in the HeteroFL \cite{b1} framework, clients are categorized into five distinct complexity levels, labeled \{a, b, c, d, e\}, predicated on a dynamic adjustment mechanism responsive to the available computational resources, guided by a hidden channel shrinkage ratio of 0.5. The model rates corresponding to these complexity levels are defined as \{1, 0.5, 0.25, 0.125, 0.625\}, respectively. This implies that clients within class 'a' possess adequate computational resources to engage in the training of the entire global model, whereas those in class 'b' are limited to training on 50\% of the global model’s parameters.

To optimize the allocation of model sizes in accordance with the diverse processing capabilities of clients, we introduce a methodology encapsulated in Algorithm \ref{alg:determineModelSize}, which aims to determine the optimal model size for a client based on their ability to process a specified number of batches.

Algorithm \ref{alg:determineModelSize} begins with an initial model size of 1, representing the full model, and iteratively halves the size (Line 1). Then the number of batches the client needs to execute is computed (Line 2). At each step, it checks if the client's batch processing capacity meets the requirements for the current model size (Line 3-8). The optimal size is returned when a client's capacity matches or exceeds the threshold. If a client's capacity is insufficient for even the smallest model size, a default size of $\mu=$0.0625 is returned, indicating eligibility for a significantly reduced model (Line 9).

This dynamic model size allocation mechanism ensures that each client participates in the federated learning process to the fullest extent permissible by their computational resources. By facilitating a resource-aware distribution of model sizes, this approach enhances the overall efficiency and scalability of the CAMA framework, optimizing resource utilization across the federated network and bolstering the efficacy of the global model training process.

\subsection{Local training and aggregation}
Based on the above client selection strategy, the server primarily selects a minimum of $n$ clients and may even select additional clients even if clients have less access to excess renewable energy and spare computing capacity. This approach poses challenges in FedZero, where only clients that can execute the minimum specified number of batches are selected. However, in CAMA, even clients with limited resources can be chosen and assigned a smaller model size.

The model partitioning and aggregation techniques are implemented as described in the HeteroFL \cite{b1} paper. Based upon the model size a client gets, the server sends only that part of the global model to the client, which further decreases the communication costs. Therefore, selecting subsets of models takes place in the following manner: \\
A layer $W_g$ in the global model has dimensions $d_g \times k_g$ representing the output and input channel sizes of this layer. Then a subset of the layer with dimensions $(d_g * m) \times (k_g * m)$ is sent to a client with model size $m$ using the ordered dropout mechanism.

The client updates its local model with the parameters shared by the global model, trains the model on its local dataset, and then sends the trained parameters back to the server. Once the server receives all the varying sized trained models from the clients it aggregates through the following method:
For each element in the global model, it performs a weighted average aggregation of that element across all local models where it is present. The weight of a local model is determined by the number of examples it has been trained on, which the client shares along with the results.

For optimization in the case of non-IID data distribution among clients, we employ the "masking trick" as defined in \cite{b1}. In this method, clients mask out labels not present in their training set. During the aggregation of the last layer elements of local models, we only consider the elements corresponding to the labels that were included in the training sets.
The track of batch normalization of local clients model is set to False to avoid privacy concerns while sharing the running mean and variance of the training data, such an adaptation of BN is known as static batch normalization(sBN). After the training process finishes, the server sequentially queries local clients and cumulatively updates global BN statistics. There exist privacy concerns about calculating global statistics cumulatively and we hope to address those issues in the future work \cite{b1}.

\section{Experiments and results}

In our study, we conducted experiments utilizing two distinct datasets: MNIST and CIFAR-10. These datasets were analyzed using two different models: a Convolutional Neural Network (Conv) and ResNet-18 \cite{resnet}. The detailed experimental settings and hyperparameters can be found in Table \ref{tbl:1}.

\begin{table}[htbp]
\centering
\caption{Experimental Settings}
\label{tab:hyperparameters}
\small
\begin{tabularx}{\columnwidth}{|>{\raggedright\arraybackslash}X>{\raggedright\arraybackslash}X|}
\hline
\textbf{Parameter} & \textbf{Value} \\
\hline
\multirow{2}{*}{Model} & MNIST: CNN \\
 & CIFAR-10: ResNet18 \\
\hline
Number of Clients & 100 \\
\hline
Maximum Fraction of Client Selection & 0.1 \\
\hline
Dirichlet Distribution Beta & 0.5 \\
\hline
labels\_per\_user & \multirow{2}{*}{2} \\
(for balanced non-iid \cite{b1}) & \\
\hline \hline
\multicolumn{2}{|l|}{\textbf{Client Selection}} \\
\hline 
Alpha & 1 \\
\hline
Exclusion Factor & 1 \\
\hline \hline
\multicolumn{2}{|l|}{\textbf{Optimizer Settings}} \\
\hline
Optimizer & SGD \\
\hline
Learning Rate & 0.001 \\
\hline
Momentum & 0.9 \\
\hline
Weight Decay & 5.00e-04 \\
\hline \hline
\multicolumn{2}{|l|}{\textbf{Power Domain Setting}} \\
\hline
\multicolumn{2}{|l|}{Solar forecast data provided by Solcast \cite{solcast}} \\
\hline
\end{tabularx}
\label{tbl:1}
\end{table}

\begin{table}[htb]
    \centering
    \caption{Cumulative Energy Usage by Round and Model for Dirichlet distribution of CIFAR10 dataset.}
    \label{tab:cumulative_energy_usage}
    \begin{tabular}{|c|c|c|c|}
        \hline
        \multicolumn{4}{|c|}{\textbf{Cumulative Energy Usage (kWh)}} \\
        \hline
        \textbf{Round} & \textbf{CAMA (BN True)} & \textbf{CAMA (BN False)} & \textbf{FedZero} \\
        \hline
        1 & 0.1451 & 0.1451 & 0.08 \\
        5 & 0.7001 & 0.6105 & 0.71 \\
        10 & 1.285 & 1.2595 & 1.56 \\
        15 & 1.9226 & 1.8197 & 2.17 \\
        \hline
    \end{tabular}
\end{table}

\begin{table*}[htb]
    \centering
    \caption{Comparison of Performance Metrics for Frameworks using Dirichlet Distribution Over 15 Rounds}
    \label{tab:performance_metrics}
    \begin{tabular}{|l|l|c|c|c|c|c|c|}
        \hline
        \textbf{Dataset} & \textbf{Model} & \textbf{Max Accuracy} & \textbf{Final Accuracy (\%)} & \textbf{Avg Accuracy (\%)} & \textbf{Accuracy Std Dev (\%)} & \textbf{Total Energy Usage (kWh)} \\
        \hline
        CIFAR-10 & CAMA(BN True) & 69.58 & 69.58 & 54.08 & 17.08 & 1.85 \\
        CIFAR-10 & CAMA (BN False) & 50.6 & 44.05 & 34.86 & 10.61 & 1.84 \\
        CIFAR-10 & FedZero & 67.04 & 67.04 & 51.07 & 15.19 & 2.16 \\
        MNIST & CAMA & 93.44 & 93.44 & 63.02 & 31.97 & 2.26 \\
        MNIST & FedZero & 91.62 & 91.62 & 53.81 & 34.78 & 2.59 \\
        \hline
    \end{tabular}
\end{table*}

\begin{table}[htb]
    \centering
    \caption{Accuracy Comparison by Round and Model, using Dirichlet Distribution on CIFAR-10}
    \label{tab:accuracy_comparison}
    \begin{tabular}{|c|c|c|c|}
        \hline
        \textbf{Round} & \textbf{CAMA (BN True)} & \textbf{CAMA (BN False)} & \textbf{FedZero} \\
        \hline
        1 & 0.263 & 0.1616 & 0.2459 \\
        5 & 0.672 & 0.281 & 0.4792 \\
        10 & 0.6837 & 0.2584 & 0.622 \\
        15 & 0.6899 & 0.4851 & 0.6704 \\
        \hline
    \end{tabular}
\end{table}

We model the power domain and setup the global scenario as defined in FedZero \cite{b3} based on real solar and solar forecast data provided by Solcast \cite{solcast}. We assume a constant power supply for steps within this period. Clients are randomly distributed over the ten power domains, with each having a maximum output of 800 W \cite{b3}.



For calculation of energy consumption, similar to the \cite{b3}, the clients are assigned one of the classes among $small$, $medium$ and $large$, which is randomly representing the hardware type it has. The classes are roughly based on the performance and energy usage characteristics of T4, V100, and A100 GPUs, respectively, which maximum consume 70W, 300W, 700W of energy \cite{b3}. 
The energy consumption depends on the following factors:
\begin{itemize}
    \item The type of hardware used by the client, which determines energy consumption per batch
    \item The number of batches executed in that round
    \item Client's model rate (model size) used in that round
\end{itemize}
Therefore, the energy consumption by client $c$ in round  $i$, $E_{c, i}$ is given by
\begin{equation}
    \begin{split}
        E_{c, i} = e_p \times b_c \times mr
    \end{split}
\end{equation}

Here, $e_p$ represents the energy consumed by a model of size 1 to execute a batch. $b_c$ denotes the number of batches executed by client $c$ in that round, which equals the product of the batches in the client’s trainloader and the number of epochs, and  $mr$ is the model size or model rate of the client's model.

In this study, we delve into the intricacies of energy consumption and model accuracy within the realm of federated learning, presenting a comprehensive analysis of our proposed framework in juxtaposition with the established baseline, FedZero \cite{b3}. Our exploration is anchored in a comparative study over 15 rounds of execution, meticulously averaging results across five iterations to ensure robustness and reliability in our findings. 

The empirical data, as showcased in Figure \ref{fig:energy_comp_dir}, underscore a nuanced understanding of energy dynamics. Compared to FedZero, our framework demonstrates a nuanced balance between energy efficiency and computational efficacy, particularly evident in the cumulative energy usage depicted in Table \ref{tab:cumulative_energy_usage}. This balance is crucial for sustainable federated learning deployments, especially in environments where energy resources are constrained or costly.

Figure \ref{fig:acc_comp_dir} presents a comparison of accuracy between our framework, with the track parameter set to True and False respectively in the batch normalization layers, and FedZero. The experiment is carried out on ResNet18\cite{resnet} model across 15 rounds on the CIFAR10 dataset distributed among clients using Dirichlet distribution. FedZero utilizes the model architecture defined in torchvision, which, by default, has batch normalization layers are set to True. Figure \ref{fig:energy_comp_dir} illustrates a comparison of energy consumption, while Table \ref{tab:accuracy_comparison} presents the accuracy comparison at specific intervals, offering insights into the convergence speeds of both frameworks under the same experimental setup. Similarly, Table \ref{tab:performance_metrics} compares the accuracies of the frameworks using the CIFAR-10 and MNIST datasets, also under the same experimental conditions.

Figure \ref{fig:acc_comp_bal_niid} depicts the comparison of accuracy on balanced non-IID distribution of CIFAR-10 dataset among the clients, where each client has at most two class labels.

\begin{figure*}[htb]
    \begin{minipage}{0.32\textwidth}
        \centering
        \includegraphics[scale=0.225]{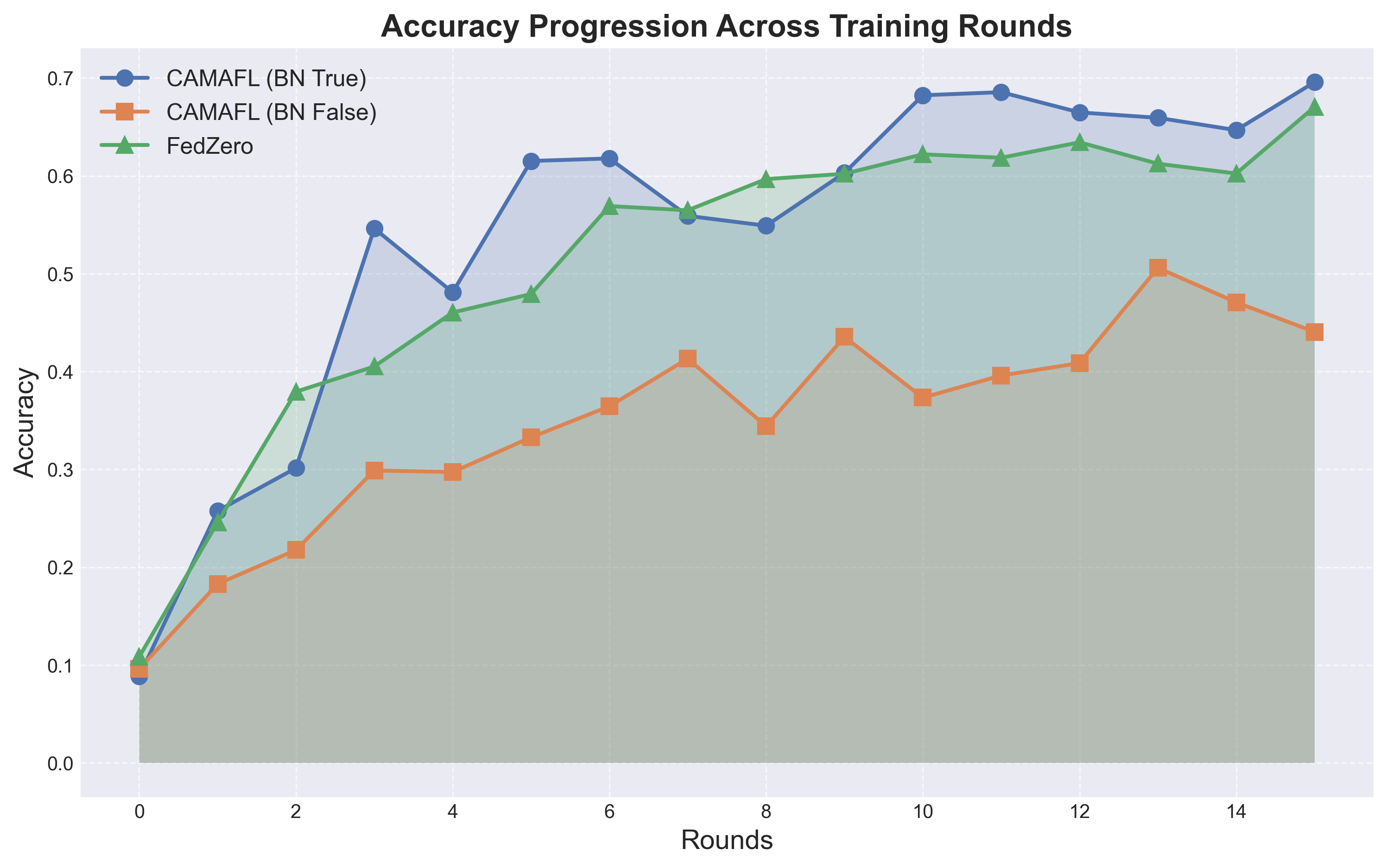}
        \caption{Accuracy for Dirichlet distribution on CIFAR10 dataset. CAMAFL with batch normalization obtains a better average performance than FedZero.}
        \label{fig:acc_comp_dir}
    \end{minipage}\hfill
    \begin{minipage}{0.32\textwidth}
        \centering
        \includegraphics[scale=0.225]{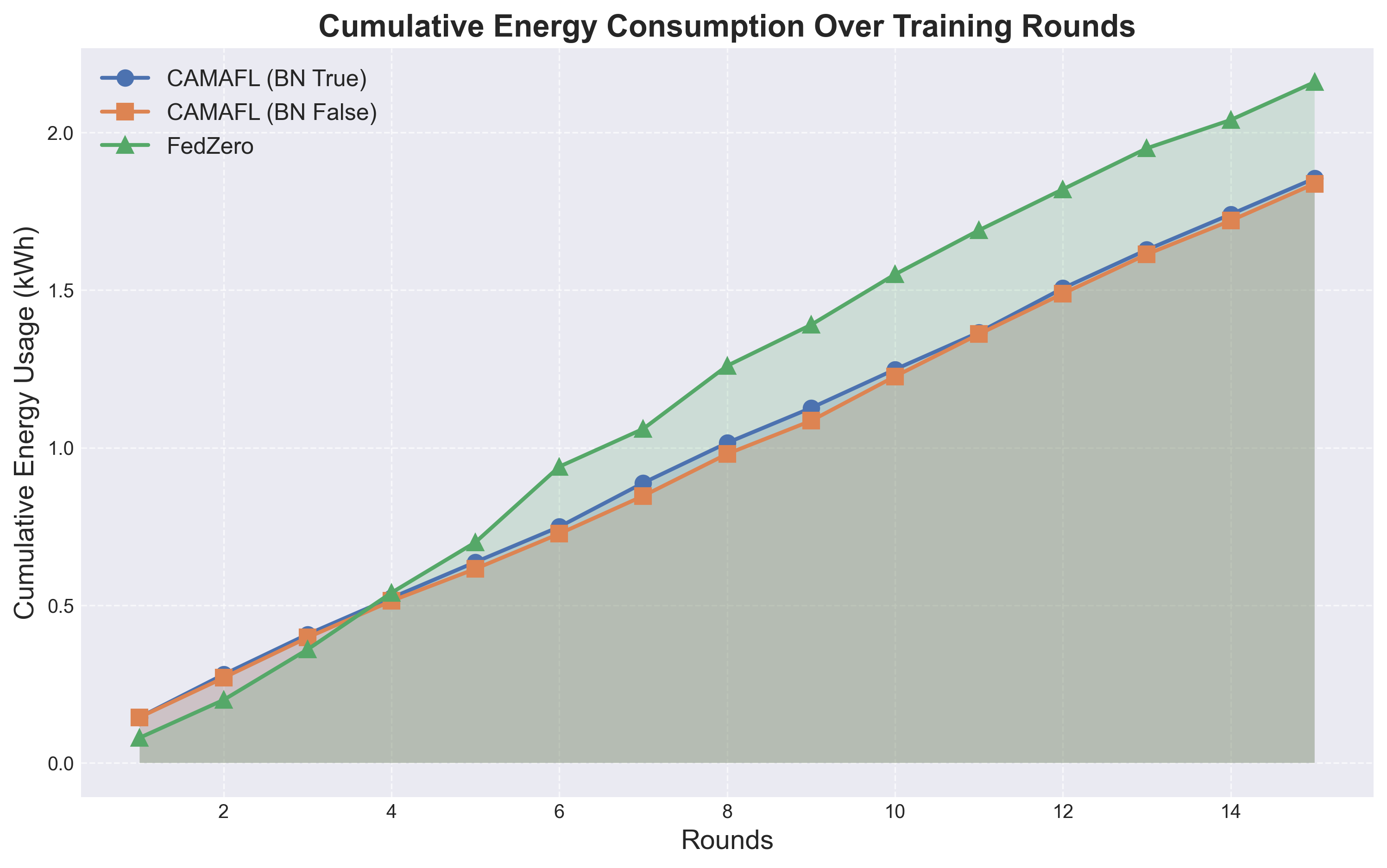}
        \caption{Energy consumption over training rounds. CAMAFL gains better average performance, with significantly lower energy consumption.}
        \label{fig:energy_comp_dir}
   \end{minipage}\hfill
   \begin{minipage}{0.32\textwidth}
        \centering
        \includegraphics[scale=0.24]{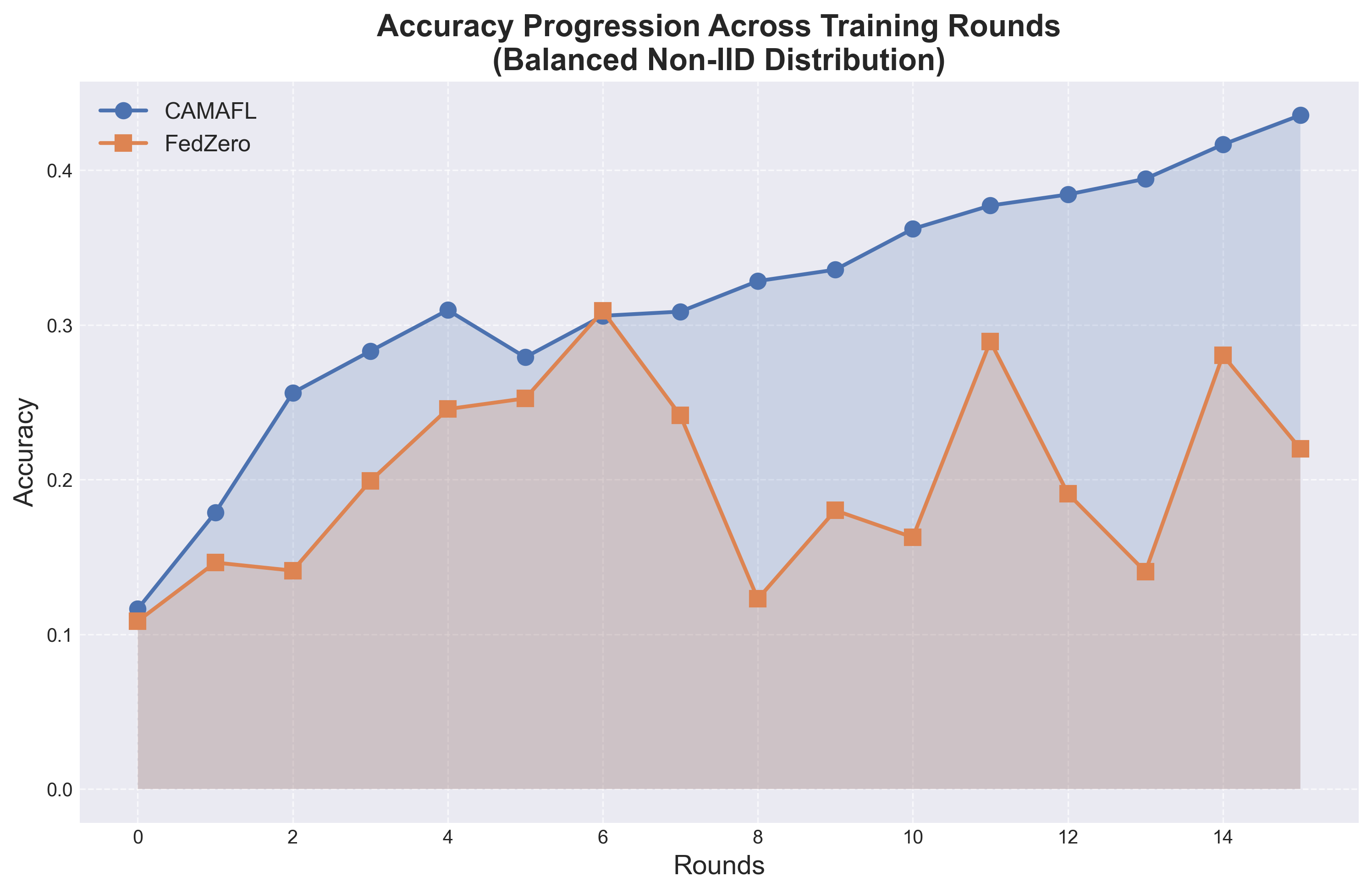}
        \caption{Accuracy for balanced non-IID distribution. CAMAFL obtains a better and more stable accuracy      progression compared to FedZero.}
        \label{fig:acc_comp_bal_niid}
    \end{minipage}\hfill
\end{figure*}

\section{Conclusion}
We introduce CAMA, a method that adjusts model sizes based on the availability of excess renewable energy and spare computing capacity. Our proposal is designed to be robust against the challenges posed by non-IID data distribution. Through extensive experiments, we have demonstrated that it consistently outperforms baseline approaches, particularly in scenarios where the dataset labels are more unevenly distributed among clients, making it highly effective in real-world federated learning environments.

\section*{Acknowledgment}
The authors would like to acknowledge the support of the Mitacs Globalink Research Internship (GRI). This research is supported in part by the Natural Sciences and Engineering Research Council of Canada (NSERC) Discovery Grant RGPIN-2024-03954.

\bibliographystyle{IEEEtranN}
\bibliography{references}

\end{document}